\documentclass{AUJarticle}
\usepackage[cmex10]{amsmath}
\usepackage[utf8x]{inputenc}
\usepackage[nocompress]{cite}
\usepackage{graphicx, multirow, booktabs, color, listings}
\usepackage{balance}
\usepackage{url}

\graphicspath{{figures/}}

\newcommand\correspondingauthor{\thanks{Corresponding author.}}

\newcommand{\squeezeup}{\vspace{-2.5mm}}

\begin{document}

\title{Reconfigurable and Scalable Honeynet for Cyber-Physical Systems}

\addauthor{Luís Sousa, José Cecílio, Pedro Ferreira, Alan Oliveira\correspondingauthor}{LASIGE, Faculdade de Ciências, Universidade de Lisboa}{fc60428@alunos.fc.ul.pt,\{jmcecilio, pmferreira, aodsa\}@ciencias.ulisboa.pt}

\issuev{35}
\issuen{1}
\issued{March 2024}

\shortauthor{L. Sousa, J. Cecílio, P. Ferreira, A. Oliveira}
\shorttitle{Re-configurable and Scalable Honeynet for Cyber-Physical Systems}

\thispagestyle{plain}

\maketitle
\begin{abstract}

Industrial Control Systems (ICS) constitute the backbone of contemporary industrial operations, ranging from modest heating, ventilation, and air conditioning systems to expansive national power grids. Given their pivotal role in critical infrastructure, there has been a concerted effort to enhance security measures and deepen our comprehension of potential cyber threats within this domain. To address these challenges, numerous implementations of Honeypots and Honeynets intended to detect and understand attacks have been employed for ICS. This approach diverges from conventional methods by focusing on making a scalable and reconfigurable honeynet for cyber-physical systems. It will also automatically generate attacks on the honeynet to test and validate it. With the development of a scalable and reconfigurable Honeynet and automatic attack generation tools, it is also expected that the system will serve as a basis for producing datasets for training algorithms for detecting and classifying attacks in cyber-physical honeynets.


\end{abstract}

\section{Introduction}

Honeypots serve as tools designed to mislead attackers by creating an illusion that they are targeting genuine infrastructure. In reality, these systems gather valuable information about the ongoing attack, enabling the enhancement of both the actual infrastructure and the honeypot. Additionally, they provide insights into the attackers' intentions and the origin of the attack\cite{9520645}. Honeynets are multiple Honeypots working together in the same system\cite{9520645} which is the intended implementation of this work. A system like a CPS is typically considered a honeynet because each component can be considered a honeypot.

Honeynets are important for Industrial Control Systems (ICS), as the impact of a cyber-attack on these systems can cause significant harm to countries and society. Examples of some of the most prolific attacks on CPS include the Stuxnet \cite{5772960}, which attacked the nuclear enrichment facilities of Iran, or the 2015 attacks on the Ukrainian power grid \cite{7752958}. Note that the difficulty of also identifying the culprits and plausible deniability of the attacks makes it attractive for governments to influence foreign opinion and cause unrest without many consequences\cite{10.1093/cybsec/tyac007}.

A ICS is a type of Cyber-physical System (CPS) that is typically composed of actuators and sensors that interact with the physical world (called the Plant), the control is composed of Programmable Logic Controllers (PLC) or Remote Terminal Units (RTU) that control the Plant and Monitoring typically done through a Human Machine Interface (HMI). This system is typically called a Supervisory Control and Data Acquisition (SCADA). To develop a CPS honeynet it is necessary to simulate all the components of a typical CPS so that it provides realistic data to fool the attacker into thinking it is a real system. For this, the use of real-time simulations to provide realistic data and information about a physical system is required. For that, there needs to be a complete simulation of the entire system from monitoring to the physical world.

In this context, this work aims to implement a software-based scalable and reconfigurable honeynet. This means that it should be able to add components and reconfigure the existing components. This work also focuses on the generation of dynamic and orchestrated attacks in the honeynet. The attack generation capability is used to validate the system. For this, the work includes the creation of a component called the Attack coordinator that using attack modelling algorithms can attack the CPS dynamically. After having a good implementation of the honeynet and the attack generator, an additional goal is to use this setup to produce a data set for Machine Learning (ML)-based intrusion detection systems (IDS) in cyber-physical honeynets.

With this, the main objectives of this work are:
\begin{enumerate}
    \item\label{obj:DevelopHoneynet} Develop a honeynet capable of being scalable and re-configurable.
    \item\label{obj:Attackgeneration} Produce an automatic attack generation tool to validate and test the honeynet.
    \item\label{obj:MLgeneration} Using the results of objectives \ref{obj:DevelopHoneynet} and \ref{obj:Attackgeneration} create datasets for ML-based IDSs in cyber-physical honeynets.
\end{enumerate}

The rest of this work is organized as follows: Section~\ref*{sec:relwork} shows some of the work related to the topics of honeynets. Section~\ref*{sec:work1} describes the proposed system design. In Section~\ref*{sec:forthcomingwork} discusses the results achieved so far and the work that has to be done. Section~\ref*{sec:conclusion} brings the conclusions. \\

\begin{table*}[t]
    \centering
    \begin{tabular}{p{2.4cm}|l|l|p{1.5cm}|p{2.7cm}|p{3cm}}
    
         \textbf{Work}    & \textbf{Year} & \textbf{Scalability} & \textbf{LoI} & \textbf{Attack Simulation} & \textbf{Physical Interaction} \\ 
         \hline
         \hline 
         Hilt et al.\cite{hilt_factoryhoneypotwp_tr_2020} & 2020 & None& High & - & Real Hardware \\
         \hline
         HoneyVP\cite{9500567} & 2021 & Limited & High & - & PLC level \\
         \hline
         HoneyPLC\cite{HONEYPLC} & 2020 & Good & Medium & - & None \\
         \hline
         Pliatsios et al.\cite{8858431} & 2019 & Limited & Medium & - & Recorded data \\
         \hline
         ICSpot\cite{9851732} & 2022 & Good & Medium & - & MiniCPS\cite{10.1145/2808705.2808715} \\
         \hline
         PLCHoney\cite{10190640} & 2023 & Limited & High & Data injection & Matlab Simulation \\
         \hline
         MimePot\cite{8913891} & 2019 & Limited & High & Integrity Attack & Python Simulation \\
         \hline
    \end{tabular}
    \caption{Table representing the honeypots analyzed (LoI: Level of Interaction)}
    \label{tab:my_label}
\end{table*}

\section{Related Work} \label{sec:relwork}

The related works focus on similar implementations of honeynets to the honeynet this work is going to implement.

Hilt et al.\cite{hilt_factoryhoneypotwp_tr_2020} goes into detail on the implementation of a honeynet using real hardware in a real factory. Hilts implementation focuses on having a realistic system so that an attacker will be fooled, but it also shows the costs that are required for a honeynet with real hardware, and that's why this work implementation diverges from having real hardware and instead uses a simulated environment.

HoneyVP \cite{9500567} is a honeypot that uses a real PLC and routes the attack traffic to it. This approach is capable of handling multiple requests and attacks simultaneously by sending them to the same PLC. It uses Memoization\cite{Michie1968-at} to reduce the required interactions with the PLC, this implementation reduces the cost compared to a traditional full hardware implementation but doesn't offer much realism when looking at the complete environment of a CPS as it only targets the PLC. 

HoneyPLC \cite{HONEYPLC} is a honeynet for simulating PLCs. It starts with real physical PLCs and tries to copy all relevant information from the PLC. After this HoneyPLC can simulate the different PLCs that it reads allowing for scalability and diversity in what an attacker sees. This work limits its scope to the PLCs thus still lacking when it comes to the other components of a CPS.

The study \cite{8858431} (Pliatsios et al.) specifies a procedure to create a simple honeynet from an already existing CPS. It uses Conpot\cite{conpot} a honeypot framework and uses real traffic from an RTU with two sensors. Afterwards, Conpot uses that recorded traffic to generate fake traffic by replaying it. From the attacker's perspective, there seems to be real traffic thus looking like a real system.

In \cite{9851732} the authors build on top of HoneyPLC to add a physical component. In this article, they add MiniCPS\cite{10.1145/2808705.2808715} which is a CPS simulation. The use of a physical simulation improves the realism of the system compared to a system without it.

The work presented in \cite{10190640} creates a honeypot called PLCHoney. It uses OpenPLC\cite{OpenPLC} and uses data from a recorded simulink simulation to generate the Plant to PLC traffic. This implementation reduces the cost of running the simulation but by only using the data from the simulation the environment the honeypot tries to simulate is less dynamic.

MimePot is the honeypot conceived in \cite{8913891}. It uses a real-time simulation for the Plant that feeds data to an HMI to form its CPS. This CPS is simple as it only includes a PLC and an HMI but is similar to the work we intend to make by having a real-time simulation for the plant.

The implementation that this work aims to achieve diverges from the other by having good scalability, a high level of interaction, attack simulation and a simulation for the physical interactions. Some of the works focus on different aspects similar to this work from the scalability or simulation of the plant, but they either only focus on a single component of a CPS or they don't achieve real scalability. This work will also focus on the attack simulation that PLCHoney and MimePot have but will try to have multiple attacks working together instead of just one.

\section{System Design} \label{sec:work1}

\begin{figure}
    \begin{center}
    \includegraphics[width=\columnwidth]{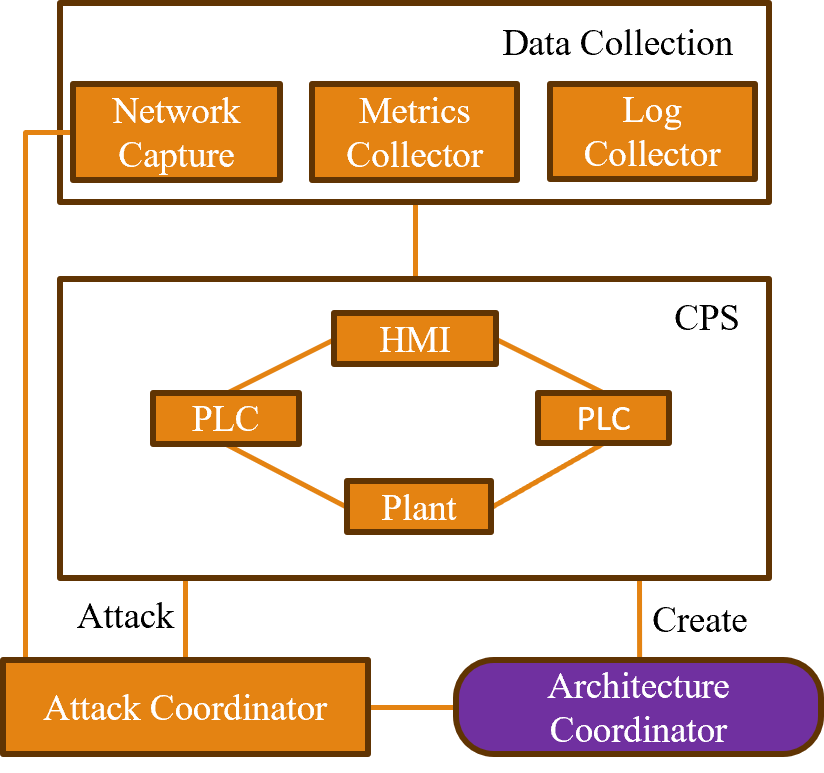}
    \end{center}
    \caption{Architecture}
    \label{fig:architecture}
    \squeezeup
\end{figure}

Figure~\ref*{fig:architecture} shows the architecture of the system. It's divided into four modules: Coordinators, Attacks, CPS and Data Collection.
The Coordinators are responsible for managing the system. This module is composed of the Architecture Coordinator and the Attack Coordinator. They communicate among themselves so that when the Architecture Coordinator generates the structure of the system the Attack Coordinator knows it as well. 

There is the CPS composed of the elements that the attacker will have access to. They are a mix of HMIs, PLCs and Plants. The number of components in this group can vary as the architecture is scalable and controlled by the Architecture Coordinator. They are networked between them in a realistic way that is meant to be similar to implementations of CPS in the industry.

Then there is the Data Collection module which is composed of the tools necessary to collect metrics, logs and network data. It will interact with all the other components so they can garner the information necessary to validate and later generate the dataset from that data.

In the implementation, all the components will be docker\cite{Docker} containers as this will help isolate every running component and make their implementation dynamic to expand and replace components. This will also help the implementation of attacks, as attacks can be docker components that initialize in runtime.

\subsection{Architecture Coordinator} \label{subsec:ArchCoordinator}

The Architecture Coordinator will only run in the setup phase so it can from a configuration file generate the necessary extra files to configure each component on the architecture and provide the required information of the architecture to the Attack Coordinator. With this information, the Attack Coordinator can orchestrate attacks that make sense on the instantiated network.

\subsection{Attack Coordinator} \label{subsec:AttackCoordinator}

The Attack Coordinator will use the architecture of the system to plan and organize attacks, this should be done dynamically for any structure of the system. This helps improve the diversity of attacks that are performed. Having diverse attacks will lead to a higher level of confidence when validating this solution. With the improved variability, we also have a more diverse and complete input to feed into a dataset.

To create realistic attacks, the attack coordinator needs to have a good idea of the architecture of the system and for this, it needs to keep track of all the useful system information. The information of the network is in a graph containing the information of each service running in each component and the possible corresponding attacks for each service.

The coordination of multiple attacks will require information about what each attack does and what changes they do to the system. This is important to make a good attack sequence where the attacks need to make changes to the system to achieve the attacker's goal.

The attacks were chosen from \cite{7924372} which specifies common attacks in CPSs. From there, the following attacks were chosen to be implemented based on how common and the complexity of implementation:
\begin{itemize}
    \setlength\itemsep{-0.8em}
  \item Man-in-the-middle
  \item Modbus-Register Reading
  \item Denial of service
  \item Modbus-Register spoofing
  \item Replay attack
\end{itemize}

More attacks can be added as we progress with the implementation.

\subsection{CPS}

This section describes the components of the CPS module. The Plant is described in the Section~\ref*{subsec:Plant}, the SCADA/HMI in Section~\ref*{subsec:SCADA}, and the PLC in Section~\ref*{subsec:PLC}.

\subsubsection{Plant} \label{subsec:Plant}

\begin{figure}
    \begin{center}
    \includegraphics[width=0.8\columnwidth]{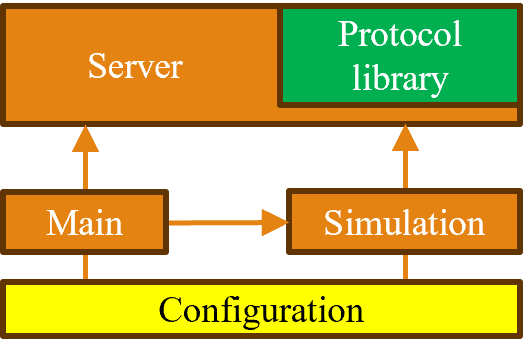}
    \end{center}
    \caption{Plant}
    \label{fig:Plant}
    \squeezeup
\end{figure}

The Plant is implemented as a container simulating the physical process of the system and interacts and communicates with the PLCs using different protocols, currently implemented Modbus/TCP. The details of the software-based Plant simulator are presented in Figure~\ref*{fig:Plant}, which represents an overview of how it works.
A simulation feeds the data to the server and can be configured with different parameters from a configuration file. The aim is to include multiple simulations and be able to choose between the different simulations with the configuration file. Currently, the implemented simulation is the heating up of a gas container following Boyle's law.

\subsubsection{SCADA/HMI} \label{subsec:SCADA}

The SCADA is implemented with an open-source solution. The one used is FUXA which is a simple HMI that is easy to configure and use.

This solution also allows some further expansion. By allowing more protocols that can add additional attacks making a more diverse and realistic environment, the diverse protocols also serve to create background network traffic.

\subsubsection{PLC} \label{subsec:PLC}

For the PLC we used OpenPLC which is also an open-source implementation of a PLC that implements our used protocols. In this case, it can use Modbus to communicate with the Plant and with the SCADA.

This solution will allow us to control the Plant with a realistic control structure. The code implemented on the PLC for the simulation follows the simple instructions of heating the container if the pressure and the temperature are below the specified on the HMI.

\subsection{Data collection}

To collect the data, the network traffic of all the components running on containers will be collected and run through a pre-processing step that will add information from the type of data in the packet.

\begin{figure}
    \begin{center}
    \includegraphics[width=0.85\columnwidth]{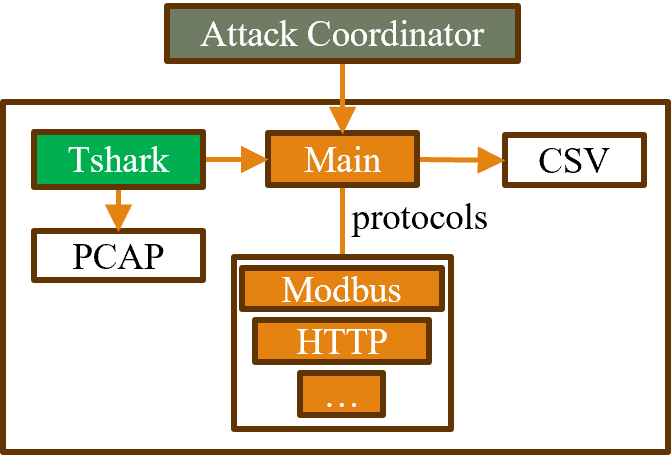}
    \end{center}
    \caption{Network Capture}
    \label{fig:NetworkCapture}
    \squeezeup
\end{figure}

The network capture will be implemented to allow the Attack Coordinator to inform which packets are attacks, and add information to the traffic about what is passing through the network. Beyond the data about the attacks, there will also be some of the data added about the kinds of protocols being used as can be seen in Figure~\ref*{fig:NetworkCapture}. 

There will be data collected beyond the network traffic. The system will also collect the metrics of the running system, from the number of network connections to processes running on the containers. The collection of logs will also be important and can be useful in feeding into the dataset to train a machine-learning algorithm, to fulfil Objetive~\ref*{obj:MLgeneration}.

\section{Current Status and Forthcoming Work} \label{sec:forthcomingwork}
Currently the Plant, PLC and HMI are implemented in the CPS thus being able to run a simple CPS. The initial work in the Architecture Coordinator and Network Capture is being done. After the completion of those two components the focus will be in the Attack Coordinator to start the realization of attacks on the system to validate and generate the data-set.


\section{Conclusion} \label{sec:conclusion}
This work aims to improve the current state of the art in honeynets by creating a honeynet that can fit any ICS, it also tries to provide better training data to create better IDSs so they can be integrated with machine learning and artificial intelligence by providing a controlled environment to generate and execute attacks on a system that can follow the structure of a real CPS.


\section{Acknowledgement} \label{sec:ack}

This work was supported by FCT through the LASIGE Research Unit, ref.\ UIDB/00408/2020 (https://doi.org/ 10.54499/UIDB/00408/2020) and ref.\ UIDP/00408/2020 (https://doi.org/10.54499/UIDP/00408/2020).

\bibliographystyle{ieeetr}
\bibliography{bibliography}
\balance

\end{document}